\documentclass[12pt]{article}
\usepackage{epsfig,a4wide}

\begin{document}

\setlength{\parskip}{0.45cm}
\setlength{\baselineskip}{0.75cm}

\begin{titlepage}
\setlength{\parskip}{0.25cm}
\setlength{\baselineskip}{0.25cm}
\begin{flushright}
DO-TH 01/14\\
\vspace{0.2cm}
hep--ph/0109284\\
\vspace{0.2cm}
\end{flushright}

\vspace{1.0cm}
\begin{center}
\LARGE
{\bf Reliability of $K\to\pi\pi$ matrix element calculations}
\footnote{Invited talk presented at the International Conference on CP
Violation: KAON 2001 (June 12--17, 2001; Pisa, Italy)}

\vspace{1.5cm}
\large
E.A.\ Paschos\\
\vspace{1.0cm}

\normalsize
{\it Universit\"{a}t Dortmund, Institut f\"{u}r Physik,}\\
{\it D-44221 Dortmund, Germany} \\
\end{center}

\begin{abstract}
\noindent
I demonstrate that the short distance contribution to 
$K\to\pi\pi$ decays must be supplemented with large distance
effects.  A hybrid calculation is outlined based on QCD
diagrams supplemented by chiral contributions and
 $\pi$--$\pi$ phaseshifts.
\end{abstract}

\end{titlepage}

\noindent 
At this meeting the KTeV group reported a new and smaller
value of $\varepsilon'/\varepsilon$.
  The experimental  values of the two groups are now
consistent with each other 
\begin{eqnarray}
\varepsilon'/\varepsilon & = &
 (15.3\pm 2.6)\times 10^{-4}\quad\quad\quad\quad{\rm{NA48}}\,\,\,
 {\rm{\cite{ref1}}}\\
\varepsilon'/\varepsilon  & = &
 (20.7\pm 2.8) \times 10^{-4}\quad\quad\quad\quad{\rm{KTeV}}
 \,\,\,{\rm{\cite{ref2}}}\, .
\end{eqnarray}
On the theoretical side there were, for a long time, two types
of predictions:\\
i) those which predicted small, almost unmeasurable, 
or negative values, and\\
ii) those which predicted large, positive and observable
values.

\noindent It is interesting to examine the reasons that led to the diverse
predictions, and also the changes required in order to bring them
into agreement with the experiment.  For my part, since I belong to
the second group,
I will present our efforts and work that led to large values
for the ratio.  The starting point of the analysis is the 
effective $\Delta S=1$ Hamiltonian derived from QCD.  The
Wilson coefficients in the Hamiltonian have been calculated to
NLO and I shall use them.  
It has become evident that tree diagram values for the
matrix elements of operators are not sufficient and we need 
to include
loop corrections, which are in fact unitarity corrections.
 The study and calculation of the loops
modify the values for the amplitudes and improve the accuracy 
of the predictions.
All groups which included loop corrections to matrix elements
predicted large 
and positive values for $(\varepsilon'/\varepsilon)$.  Now that this 
fact has been recognized
we face the next problem of estimating the uncertainties and improving the 
reliability of the results.  This is a difficult problem
because it requires defining a low energy limit of QCD, including
loops.

The first article to include loop corrections to matrix 
elements  
\cite{ref3} studied the 
operators $Q_1$ and $Q_2$ and found an enhancement of the
$I=0$ amplitude.  Subsequently, it was recognized that
loops are also important for the operators $Q_6$ and $Q_8$
\cite{ref4}, because they enhance $\langle Q_6\rangle_0$ and
reduce $\langle Q_8\rangle_2$ \cite{ref4}, thus eliminating 
the 
cancellation between the two terms and predicting a large
ratio
$(\varepsilon'/\varepsilon)$.  Loop corrections have also been
calculated in the chiral quark model \cite{ref5} with similar
results.  Finally, they were calculated in chiral
perturbation theory, paying special attention to the matching
of the chiral loops to the QCD--scale \cite{ref6} residing 
in the Wilson coefficients. 

The fact that loop corrections are important is
well accepted now.  The remaining issue is the accuracy of
such calculations, especially the dependence of the results
on the cut--off of the chiral theory, since chiral theory
is not renormalizable.  It was shown in article \cite{ref6}
that to
$O(p^2)$ the infinities of the factorizable diagrams
are absorbed into the renormalized coupling constants
\begin{displaymath}
L_5^r=2.07\times 10^{-3}\quad\quad{\rm{and}}\quad\quad
 L_8^r= 1.09\times 10^{-3}\, .
\end{displaymath}
This happens for all operators -- those that include 
currents, as
well as those with densities of quarks. The 
non--factorizable terms bring to $O(p^2)$ a quadratic
dependence on the cut--off, which makes the predictions
sensitive functions of the cut--off.  In the meanwhile,
the experience gained in the calculations, together with
the fact that we must account for five physical quantities
$|A_0(m_K)|$, $|A_2(m_K)|$, $\delta_0^0(m_K)$, 
$\delta_0^2(m_K)$ and $(\varepsilon'/\varepsilon)_K$, 
allow me to abstract four general properties that must be included in
explanations of the quantities. The aim here
is to arrive at a low energy realization of QCD which
reproduces current algebra at the 1-loop level, since current
algebra is already satisfied at the tree level of chiral
theory.

I summarize the properties as four benchmarks.\\
{\underline{Benchmark 1}}.  The imaginary parts of the
amplitude can be calculated by unitarity to be
\begin{equation}
Im\, A(K^+\to\pi^+\pi^0)=|A(K^+\to\pi^+\pi^0)|
 \left(-\frac{1}{8}\right) \frac{m_K^2-m_{\pi}^2}{4\pi F_{\pi}^2}
  \sqrt{1-\frac{4m_{\pi}^2}{m_K^2}}
\end{equation}
and similarly
\begin{equation}
Im\, A_0(K\to\pi\pi)=|A_0(K\to\pi\pi)|\,\,\frac{1}{4}\,\,
  \frac{m_K^2-0.5m_{\pi}^2}{4\pi F_{\pi}^2}
   \sqrt{1-\frac{4m_{\pi}^2}{m_K^2}}
\end{equation}
where $|A(K^+\to\pi^+\pi^0)|$ and $|A_0(K\to\pi\pi)|$ are
absolute values of the complete decay amplitudes for 
$K^+\to \pi^+\pi^0$ and the $I=0$ decay, respectively.
Taking the ratios $Im\, A/|A|$, one arrives at
\begin{eqnarray}
\sin\delta_0^0 & = & \frac{1}{4}\,\,
  \frac{m_K^2-0.5m_{\pi}^2}{4\pi F_{\pi}^2}
  \sqrt{1-\frac{4m_{\pi}^2}{m_K^2}} 
    = 0.46\\
\sin\delta_0^2 & = & -\frac{1}{8}\,\,
   \frac{m_K^2-2m_{\pi}^2}{4\pi F_{\pi}^2}
   \sqrt{1-\frac{4m_{\pi}^2}{m_K^2}}
     = 0.20
\end{eqnarray}
which give the following values for the phaseshifts\\
\indent $\delta_0^0 = 27.4^o$ 
to be compared with $34.2^o$ from
experiments, and\\
\indent $\delta_0^2=-11.5^o$ to be compared with $-7^o$ to 
  $-12^o$ from experiments.

\noindent One must conclude that for $s=m_K^2$ the absolute values of
$K\to\pi\pi$ amplitudes, combined with the phase space integration
(unitarity), gives acceptable phases.  It still remains to compute
the real parts of the amplitudes at $s=m_K^2$, as well as real parts
of the matrix elements for various operators. 

\noindent{\underline{Benchmark 2}.  Practically all 
articles use the effective $\Delta S=1$ Hamiltonian
derived from QCD.  At a high momentum scale, like
$\mu=2$ GeV, the Hamiltonian should give reliable
estimates which are then extrapolated to lower momenta.
In addition, at the high 
$\mu$--scale, the Wilson coefficients in the various
schemes are close to each other \cite{ref6a}.  Since the quarks
carry large momenta, I will use the lowest order 
contribution of quark operators between hadrons.  The numerical estimates
of such calculations are shown in Table 1, where  I give
numerical values for two momenta -- $\mu=2.0$ and $\mu=1.0$
GeV.  The change between the two scales  is relatively small,
so I have also included in the
table the experimental values for comparison.
The theoretical estimate for $A_0(\mu)$ gives only 30\%
of the experimental value and for $A_2(\mu)$ it gives
150\% of the experimental value.  Thus additional
(non--perturbative) corrections must be large and they 
must increase $A_0(m_K)$ and decrease $A_2(m_K)$.
The corrections found by the Dortmund \cite{ref4,ref6} and Trieste
\cite{ref5} groups,
in chiral theory, increase $\langle Q_6\rangle_0$ and
decrease $\langle Q_8\rangle_2$ in agreement with the
above requirement.  Similarly, the $\pi$--$\pi$ phase
shift for $I=0$ is positive and for $I=2$ negative,
bringing  corrections with the required sign
\cite{ref7,ref8,ref9}.

\begin{center}
\begin{tabular}{|lll|} 
\hline
$\mu$ & $A_0(\mu)$ in GeV & $A_2(\mu)$ in GeV\\
\hline
2 GeV & $0.62\times 10^{-7}$ & $0.25\times 10^{-7}$\\
1 GeV & $0.83\times 10^{-7}$ & $0.23\times 10^{-7}$\\
\hline
Exper. & $3.33\times 10^{-7}$ & $0.15\times 10^{-7}$\\
\hline
\end{tabular}
\vspace{0.3cm}

Table 1:  Contribution from QCD alone
\end{center}

\noindent{\underline{Benchmark 3}}.  It is evident from
Table 1 that the change of the scale $\mu$ is not large
enough to provide the required corrections.  In fact,
several operators of QCD have a similar development between
2.0 and 1.0 GeV. The running of $C_6(q^2)$ (dashed curve) and
$C_6(q^2)/m_s^2(q^2)$ (solid curve) in this momentum region are
plotted in figure 1. The variation of the ratio is much smaller than
the running of $C_6(q^2)$ alone.

\vspace{0.5cm}
\centerline{\epsfig{file=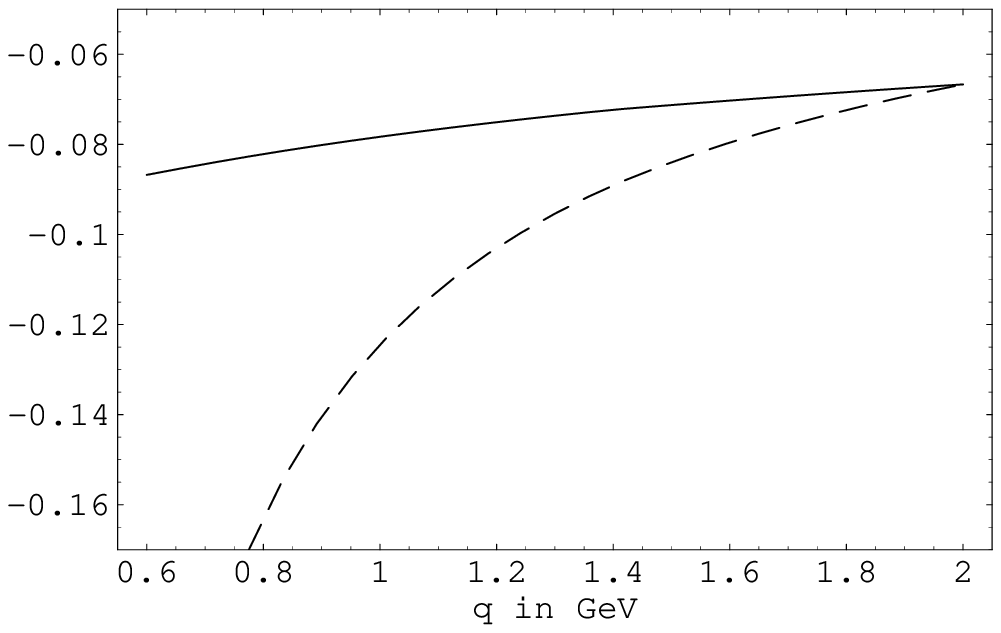,width=9cm}}
\vspace{-0.5cm}
\begin{center}
\footnotesize{Figure 1: Evolution of $C_6(q^2)$ (dashed curve) and
$C_6(q^2)/m_s^2(q^2)$ (solid curve) as functions of $q$}
\end{center}
This stability was already noticed in \cite{ref10}  
and a similar one appears for $C_8(q^2)/m_s^2(q^2)$.

Operators whose anomalous dimensions are zero are
called marginal and they remain constant over extended
regions of momenta.  A similar property appears,
numerically, in the sum of coefficients \cite{ref6a} 
\begin{equation} 
Z_1(\mu)+Z_2(\mu)\sim 0.7\,\,{\rm{to}}\,\,0.8\quad
 {\rm{for}}\quad 0.6<\mu<2.0\,\,{\rm{GeV}}\, .
\end{equation}

\noindent These results from QCD suggest that marginal
quantities of QCD could be extrapolated to low momenta.
In this case, large corrections
must reside on the values of the matrix elements as 
non--perturbative effects.

\vspace{0.3cm}
\begin{minipage}[c]{7cm}
\centerline{\epsfig{file=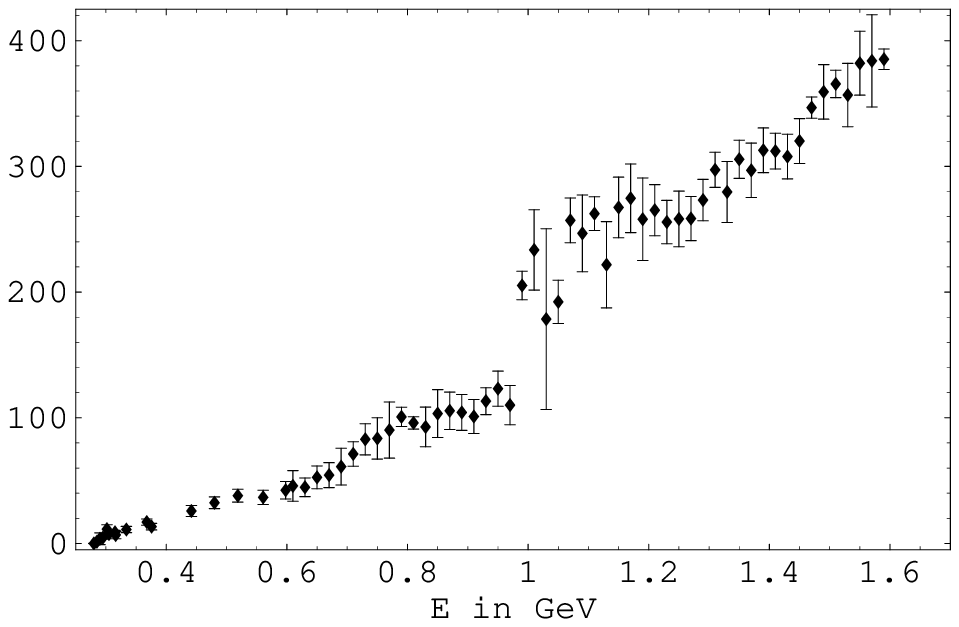,width=7.5cm}}
\begin{center}
\footnotesize{Figure 2a:  The phaseshift $\delta_0^0$ as function of
the energy ${\rm E}=\sqrt{s}$}
\end{center}
\end{minipage}
\hfill
\begin{minipage}[c]{7cm}
\centerline{\epsfig{file=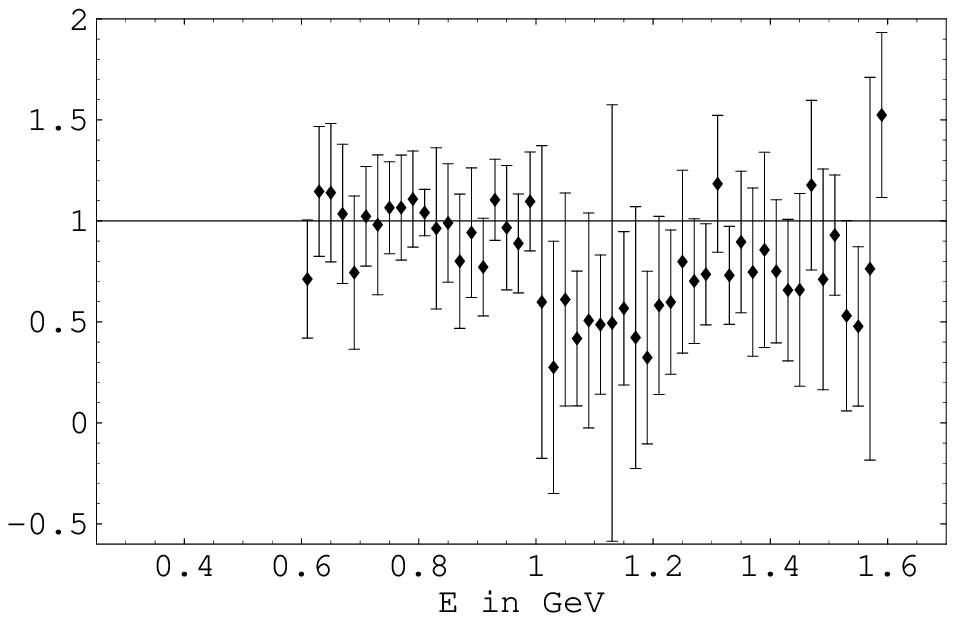,width=7.5cm}}
\begin{center}
\footnotesize{Figure 2b:  Inelasticity $\eta_0^0$ for the $I=0$ channel
as a function of the energy}
\end{center}
\end{minipage}
\vspace{1cm}

\noindent{\underline{Benchmark 4}}.  We have a lot of 
data on $\pi$--$\pi$ 
scattering, some of them being shown in \mbox{figures} 
2a and 2b.  
The data indicate that the phaseshift for
the $I=0$ channel remains elastic up to 900 MeV
and then the resonance $f_0(980)$ appears, which forces 
the $\delta_0^0$ phaseshift to pass through $180^o$, as
shown in figures 2a and 2b \cite{ref11}.
The behavior of $\delta_0^0$ must help the convergence of
the $I=0$ amplitudes, since $\sin\delta_0^0$ goes through 
zero at $E=1$ GeV.
The $I=2$ phaseshift is negative, smaller  and has no 
special structure.
The dynamics of the $\pi$--$\pi$ scattering play a role in
the decay amplitudes and must be included in the analysis.
Contributions from the phaseshifts will bring corrections
with the desired signs: increase $\langle Q_6\rangle_0$
and decrease $\langle Q_8\rangle_2$.  Suggestions along
these lines have already been discussed \cite{ref8,ref9}.

It is suggestive that an explanation of the $K$--decays must include
the above properties. There are interpolations suggested by the above
benchmarks. I will outline a method in this context, which
tries to improve the high energy region of chiral calculations ($p
\approx 1$GeV). I will replace the $K$--meson
with the divergence of the $\Delta S=1$ axial current and try to apply
current algebra. The divergence of the axial current is realized below 1.0
GeV by its chiral representation and above that energy by
the quark representation.  The amplitude is an analytic
function of the invariant mass of the two--pions--squared, $s$,
and satisfies a dispersion relation 
\begin{equation}
Re\, A_I(\sigma) = \frac{1}{\pi} \int_{4m^2}^{s_0}
\frac{{\rm Disc.}\, A_I(s)}{s-\sigma}\, ds +
 \frac{1}{\pi} \int_{s_0}^{m_t^2}
  \frac{{\rm Disc.}\, A_I(s)}{s-\sigma}\, ds\, .
\label{Re-amplitude}
\end{equation}
The value of $s_0$ is an intermediate scale, with
$\sqrt{s_0}=1.0$ to 1.4 GeV.  The splitting of the integral
is intentional so that for the high energy term I use the
QCD amplitude.  For instance, consider the amplitude
\begin{equation}
A_{Q_6}(s,\mu) = C_6(s,\mu)\langle \pi\pi|Q_6|K\rangle
\end{equation}
with the discontinuity given by \cite{ref12,ref13}
\begin{equation}
{\rm{Disc.}}\, A_{Q_6}(s,\mu) = 
 \frac{\pi}{ln^2|s/\Lambda^2|+\pi^2}\, C_6(\mu)
  \left(ln\frac{\mu^2}{\Lambda^2}\right)\, 
   \langle \pi\pi|Q_6|K\rangle\, .
\end{equation}
In this energy region the momenta in $Q_6$ are large and the tree
contribution for the matrix element will suffice.  The integral
over the discontinuity is finite and no subtraction is 
required \cite{ref13}.

The discontinuity in the low energy integral is assumed to be
\cite{ref6}
\begin{equation}
\rm{Disc.}\, A_{Q_6}(s) = -4\sqrt{3}\, L_5\, 
  \left(\frac{m_K^2-m_{\pi}^2}{F_{\pi}}\right)\,
   \frac{4m_K^4}{m_s^2(s)} \, g_0(s)
\end{equation}
with the Goldberger--Treiman factor \cite{ref14}
\begin{equation}
g_0(s) = \frac{\eta_0\sin 2\delta^0\cos\delta^0}
  {1+\eta_0\cos 2\delta^0}\, ,
\end{equation}
$\eta_0$ the inelasticity and $\delta^0$ the phaseshift.
This representation is certainly valid below the inelastic
threshold.  Since the $I=0$ channel remains elastic up to 
1.0 GeV, I
will use this form.  The enhancement factor from the first
integral in eq.\ (\ref{Re-amplitude}) is given by
\begin{equation}
F_6(s_0,m_k) = \frac{1}{\pi} \int_{4m^2}^{s_0} \,
   \frac{g_0(s)}{s-m_k^2}\,\, 
     \frac{m_s^2 (1 \, {\rm GeV} )}{m_s^2 (s) }
\left( 1 \pm h(s) \right) \, ds\, .
\end{equation}
The phenomenological function $h(s)= \frac{1}{2} \left(\frac{s-0.25}
{1.5}\right)$ in the integrand is introduced in order to study
the sensitivity of the enhancement factor on off-the-mass
shell effects \cite{ref15} (see discussion).  Figure 3 shows the enhancement
factor as a function of ${\rm E} = \sqrt{s_0}$. The factor $F_6(s_0,m_k)$
increases up to 900 MeV and then flattens out. To obtain the solid
curve I set $h(s)=0$ and the strange quark mass in the denominator
equal to $m_s$(1 GeV).  The dashed and dashed--dotted
curves include the running strange quark mass and $\pm h(s)$, respectively.
The enhancement factor lies between 0.40 and 0.55.  A similar
calculation for $F_8(s_0,m_k)$ gives a depletion factor with
the approximate value of $-0.20$ to $-0.30$.

\vspace{0.5cm}
\centerline{\epsfig{file=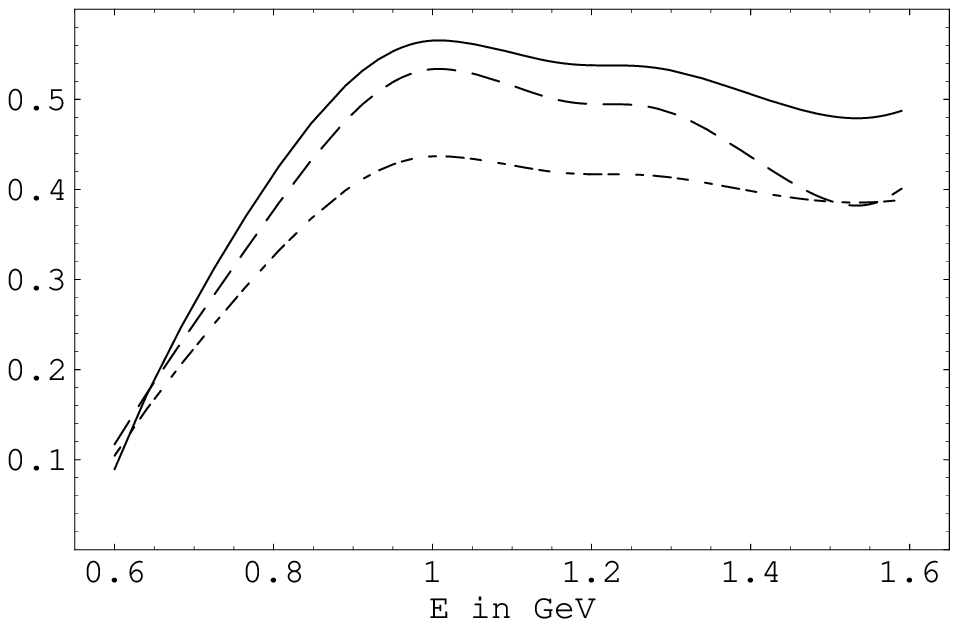,width=9cm}}
\vspace{-0.5cm}
\begin{center}
\footnotesize{Figure 3: The enhancement factor $F_6(s_0,m_k)$ as a function
of ${\rm E}=\sqrt{s_0}$}
\end{center}

The loop corrections increase $\langle Q_6\rangle_0$
by 50\% and  decrease $\langle Q_8\rangle_2$ by 20\% to 30\%.
For these values the ratio attains values
\begin{equation}
  (\varepsilon'/\varepsilon) \approx (15 \ldots 20) \times 10^{-4}\, ,
\label{theo-eps}
\end{equation}
which do not include the isospin breaking term yet. The range given in
eq.\ (\ref{theo-eps}) includes a rough estimate of the theoretical
uncertainties. A detailed calculation with a description of the
current algebra and a precise estimate of the theoretical
uncertainties will be presented in a publication.

\centerline{\large Acknowledgements}

Support of the ``Bundesministerium f\"ur Bildung, Wissenschaft,
Forschung und Technologie'', Bonn under contract 05HT1PEA9 is
gratefully acknowledged. I wish to thank Mr. V.\ Krey for discussions
and help with numerical calculations. 


\end{document}